\documentclass[proof]{WileyASNA-v1}

\articletype{Article Type}

\received{0000}
\revised{0000}
\accepted{0000}

\newif\ifmarkchanges \markchangestrue

\markchangesfalse

\definecolor{refA}{RGB}{0,80,200}
\definecolor{refB}{RGB}{205,25,25}
\definecolor{refC}{RGB}{0,135,55}
\definecolor{refD}{RGB}{225,120,0}
\definecolor{refE}{RGB}{125,45,175}
\definecolor{refF}{RGB}{0,150,170}
\definecolor{refG}{RGB}{140,85,35}
\ifmarkchanges
  \DeclareRobustCommand{\RA}[1]{\textcolor{refA}{#1}}
  \DeclareRobustCommand{\RB}[1]{\textcolor{refB}{#1}}
  \DeclareRobustCommand{\RC}[1]{\textcolor{refC}{#1}}
  \DeclareRobustCommand{\RD}[1]{\textcolor{refD}{#1}}
  \DeclareRobustCommand{\RE}[1]{\textcolor{refE}{#1}}
  \DeclareRobustCommand{\RF}[1]{\textcolor{refF}{#1}}
  \DeclareRobustCommand{\RG}[1]{\textcolor{refG}{#1}}
\else
  \DeclareRobustCommand{\RA}[1]{#1}\DeclareRobustCommand{\RB}[1]{#1}
  \DeclareRobustCommand{\RC}[1]{#1}\DeclareRobustCommand{\RD}[1]{#1}
  \DeclareRobustCommand{\RE}[1]{#1}\DeclareRobustCommand{\RF}[1]{#1}
  \DeclareRobustCommand{\RG}[1]{#1}
\fi

\makeatletter
\@ifpackageloaded{ulem}{\normalem}{}
\makeatother

\begin{document}

\title{Flavor-dependent quark-mass scaling and the fate of the strange-matter
window in self-bound quark stars}

\author[1]{G. Teruya*}
\author[1]{G. Lugones}
\author[2,3]{A. G. Grunfeld}
\authormark{Teruya, Lugones and Grunfeld}

\address[1]{\orgdiv{Centro de Ci\^encias Naturais e Humanas}, \orgname{Universidade Federal do ABC}, \orgaddress{Avenida dos Estados 5001- Bang\'u, CEP 09210-580, Santo Andr\'e \state{SP}, \country{Brazil}}}

\address[2]{\orgname{Instituto de Astronom\'{\i}a y F\'{\i}sica del Espacio (IAFE, CONICET-UBA)}, \orgaddress{Ciudad Universitaria (1428), \state{Ciudad Autónoma de Buenos Aires}, \country{Argentina}}}

\address[3]{\orgname{CONICET}, \orgaddress{Godoy Cruz 2290 (C1425FQB) CABA,   \country{Argentina} }}

\corres{*Avenida dos Estados 5001- Bang\'u, CEP 09210-580, Santo Andr\'e SP, Brazil. \email{giulliano.teruya@aluno.ufabc.edu.br}}

\abstract{In the flavor-dependent QMDD model, self-bound $ud$ quark matter can
be the ground state at vanishing pressure and convert into $uds$ matter through
a first-order transition. So far the mass functional has used parameters
$(a,C)$ common to all flavors; here we let the strange sector have its own
coupling, $C_s=\alpha\,C_{u,d}$, and map the response of the zero-pressure
parameter space. The $e^0_{ud}=930\,$MeV boundary is independent of $\alpha$ by
construction, whereas increasing $\alpha$ enlarges the self-bound $ud$ domain
and pushes the $e^0_{ud}=e^0_{uds}$ line towards larger $a$. The causality
boundary $c^0_s=c$ is almost frozen near $a\simeq5.5$. The competition closes
the strange-matter window at \RA{$\alpha_{\rm crit}\simeq1.136$}: above it every
causal self-bound parametrization is $ud$ at zero pressure, so a first-order
$ud\!\to\!uds$ transition must occur at higher pressure, though not necessarily
inside a star, since the maximum mass may be reached below the transition
density. The transition pressure and the energy-density jump grow
monotonically with $\alpha$, both scale \RB{approximately} with the available-volume fraction
$q=1-\kappa$, and the Seidov onset-instability diagnostic stays below unity.}


\maketitle

\section{Introduction}

Compact stars may host deconfined quark matter, either as a core inside a
hadronic envelope or, if a self-bound quark phase exists, as the whole star.
The classical realization of the second possibility is strange quark matter
(SQM)\RE{: the hypothesis goes back to \citet{Itoh:1970uw},
\citet{Bodmer:1971we}, and \citet{Witten:1984rs}, with a systematic stability
analysis by \citet{Farhi:1984qu}}. It has been pointed out, however, that
\emph{two}-flavor quark matter (udQM) can also be absolutely stable in bulk
without endangering the stability of ordinary nuclei, because surface and
curvature terms penalize small quark drops
\RE{\citep{Holdom:2017gdc,Madsen:1998uh,Lugones:2013ema,Lugones:2020qll}}.

Since first-principles QCD is not available at the relevant densities, this
question has to be addressed with effective descriptions. We use the
quark-mass density-dependent (QMDD) framework
\RE{\citep{Lugones:2022upj,Lugones:2023zfd,Lugones:2024ryz}}, in which
interactions are encoded in
density-dependent quasiparticle masses; thermodynamic consistency then generates
a confining, bag-like contribution to the pressure, while the masses approach
their current values at high density, mimicking chiral restoration. In its
\emph{flavor-dependent} version each mass tracks the density of its own flavor,
$M_i=M_i(n_i)$, which opens a broad domain in which strangeness is suppressed at
low pressure while the light flavors are not, so that $ud$ matter is self-bound
at $p\to0$ and the onset of $s$ quarks is delayed to finite pressure. In a
recent work~\citep{Teruya:2025tpx} we showed that this onset is not gradual: it
proceeds through a genuine first-order $ud\!\to\!uds$ transition that emerges
dynamically from the mass functional \RD{(it follows from the crossing of the
two equilibrium branches)}, giving rise to \emph{self-bound hybrid stars} with a $uds$ core
inside a $ud$ mantle.

That analysis, however, adopted a minimal parametrization in which the mass
functional
\begin{equation}
M_i = m_i + \frac{C}{n_i^{a/3}} ,
\label{eq:mass_blind}
\end{equation}
uses the \emph{same} $(a,C)$ for the three flavors, so that all the flavor
dependence comes from the current masses $m_i$ and from $\beta$ equilibrium.
Allowing an explicit flavor dependence of the parameters was left as a natural
next step. The present contribution takes that step \RD{by introducing one
additional parameter} and asks a specific question: \emph{how does the zero-pressure parameter space
respond when the strange sector is given its own coupling?} As we show below,
the answer is not a mere quantitative shift. The strange-matter window closes
altogether above a critical value of the new parameter, and beyond that point
the flavor-dependent QMDD model describes only self-bound two-flavor matter
undergoing a first-order transition. The scope of the analysis is deliberately
limited to bulk matter: the parameter space is charted at zero pressure, and
the properties of the $ud\!\to\!uds$ transition itself --- transition pressure,
energy-density jump, and the Seidov onset-instability diagnostic --- are then
examined at the finite pressure at which the conversion takes place. No stellar
structure calculation is attempted here.

\section{The model with $C_s=\alpha\,C_{u,d}$}
\label{sec:model}

We describe cold quark matter as a gas of quasiparticles of flavors $u,d,s$
with density-dependent effective masses, supplemented by free electrons. We
keep the functional form of Eq.~\eqref{eq:mass_blind} but let the strange
sector have its own coupling,
\begin{align}
M_{u,d} &= m_{u,d} + \frac{C_{u,d}}{n_{u,d}^{a/3}}, &
M_{s} &= m_{s} + \frac{\alpha\,C_{u,d}}{n_{s}^{a/3}},
\label{eq:mass_alpha}
\end{align}
with $\alpha\equiv C_s/C_{u,d}>0$ and a common exponent $a$. The case $\alpha=1$
reproduces the model of~\cite{Teruya:2025tpx}. Physically, $\alpha>1$ makes
the medium contribution to $M_s$ stronger, suppressing strangeness further at
low density, while $\alpha<1$ favors it. We use $m_u=m_d=3.45\,$MeV and
$m_s=93\,$MeV; $M_i$ is in MeV when $n_i$ is in fm$^{-3}$, and we work in
natural units.

At $T=0$ \RD{the energy density retains the free-gas functional form, while
the pressure and the chemical potentials acquire additional terms from the
density dependence of the masses,}
\begin{eqnarray}
\epsilon &=& \sum_i g\,M_i^4\,\chi(x_i)+\epsilon_e, \\
p &=& \sum_i \left[g\,M_i^4\,\phi(x_i)-B_i\right]+p_e, \\
\mu_i &=& M_i\sqrt{x_i^2+1}-\frac{B_i}{n_i},
\end{eqnarray}
where $g=6$, $x_i=(6\pi^2 n_i/g)^{1/3}/M_i$, the functions $\chi,\phi,\beta$ are
the standard free-Fermi-gas integrals, and
\begin{equation}
-B_i = g\,\beta(x_i)\,M_i^3\,n_i\,\frac{\partial M_i}{\partial n_i}
     = -\tfrac{a}{3}\,g\,\beta(x_i)\,M_i^3\,(M_i-m_i)
\end{equation}
is the effective bag term generated by the density dependence of the masses.
Short-range repulsion is included through the excluded-volume prescription of
\RE{\citet{Lugones:2023zfd,Lugones:2024ryz}}, with $b_i=\kappa/n_i$, so that the
available-volume fraction $q=1-\kappa$ is flavor independent: the quark-sector
quantities are evaluated at rescaled densities $\tilde n_i=n_i/q$ and multiplied
by $q$, the electron sector being unchanged. Explicit expressions are collected
in the Appendix of \RE{\citet{Teruya:2025tpx}}. Throughout we impose local charge
neutrality and $\beta$ equilibrium, $\mu_d=\mu_u+\mu_e$ and $\mu_s=\mu_d$.

Two quantities organize the discussion. The first is the energy per baryon at
zero pressure,
\begin{equation}
e^0 \equiv \left.\frac{\epsilon}{n_B}\right|_{p=0},
\end{equation}
evaluated separately for two-flavor ($e^0_{ud}$) and three-flavor ($e^0_{uds}$)
matter and compared with \RF{the conventional benchmark of $930\,$MeV adopted
in stability analyses of quark matter \citep{Farhi:1984qu}}. The second is the
sound speed at vanishing pressure, $c^0_s$, which
must not exceed $c$ for the self-bound configuration to be admissible. Both are
independent of $\kappa$ \RB{for the quark sector}: in the excluded-volume
construction the Gibbs energy
per baryon at $p=0$ and the ratio $dp/d\epsilon$ in the $p\to0$ limit are
unaffected by the available-volume factor \RE{\citep{Teruya:2025tpx}}.
\RB{The electrons, which are not rescaled, leave only a weak residual dependence
on $\kappa$ (negligible for three-flavor matter and below $1\%$ in $e^0_{ud}$
for $\kappa\le0.6$). The windows shown below are computed at $\kappa=0$ and are
then} fully determined by $(a,C_{u,d},\alpha)$.

\section{The stability window as a function of $\alpha$}
\label{sec:window}

\begin{figure*}[tb]
\centering
\includegraphics[width=0.88\textwidth]{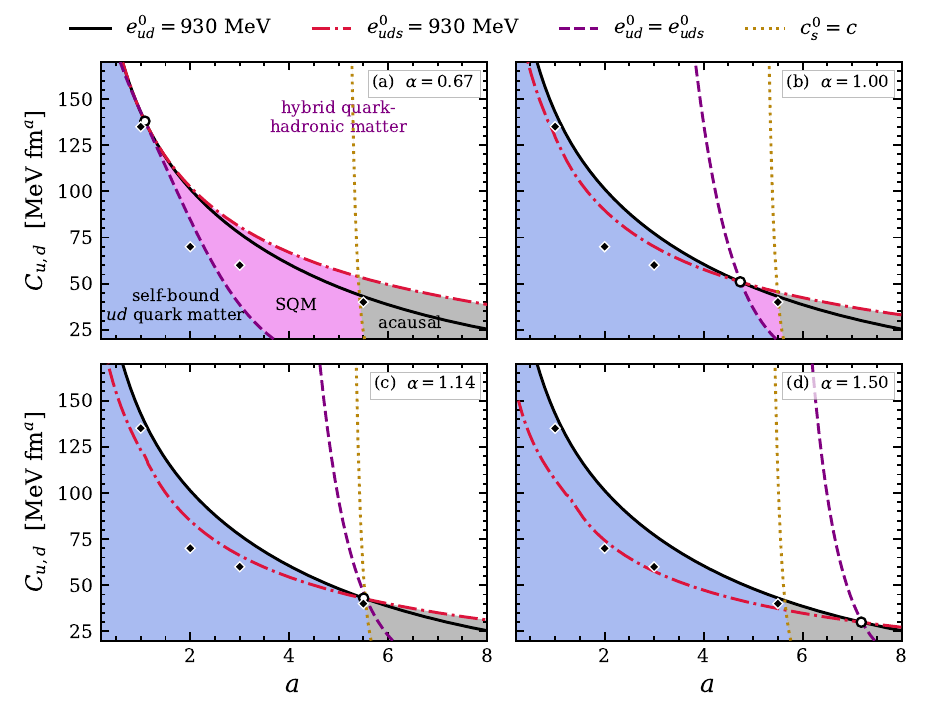}
\caption{Stability window of the flavor-dependent QMDD model in the
$(a,C_{u,d})$ plane \RB{(at $\kappa=0$)} for four values of
$\alpha=C_s/C_{u,d}$: (a) $0.67$, (b)
$1.00$, (c) $1.14$, (d) $1.50$. Black solid:
$e^0_{ud}=930\,$MeV (independent of $\alpha$); red dash-dotted:
$e^0_{uds}=930\,$MeV; purple dashed: $e^0_{ud}=e^0_{uds}$; gold dotted:
$c^0_s=c$. Blue shading marks self-bound two-flavor matter, light magenta
strange quark matter, gray self-bound but acausal parametrizations, and white
hybrid quark-hadronic equations of state.
Symbols: the open circle is the triple point $(a_{\rm tp},C_{\rm tp})$, where
the three energy boundaries meet; the black diamonds are the four benchmark
parameter sets of~\cite{Teruya:2025tpx}, $(a,C_{u,d})=(5.5,40)$, $(1,135)$,
$(2,70)$ and $(3,60)$, the last three being the ones analysed in
Sec.~\ref{sec:transition}.
The SQM domain shrinks as $\alpha$ grows and has disappeared by
$\alpha\simeq1.14$ \RA{($\alpha_{\rm crit}\simeq1.136$)}; correspondingly, the benchmark $(3,60)$ lies inside the SQM
domain for $\alpha=0.67$ but in the self-bound $ud$ domain for all larger
$\alpha$, and the strange-quark-matter benchmark $(5.5,40)$ has left the SQM
domain altogether by $\alpha=1.14$. In panel (d) the $e^0_{uds}=930\,$MeV curve
stops at $C_{u,d}\simeq150$: above that value the three-flavor sequence
terminates at positive pressure and has no self-bound point at all.}
\label{fig:1}
\end{figure*}

By stability window we mean the region of the parameter space in which quark
matter is absolutely stable at zero pressure, that is, in which its energy per
baryon lies below the \RF{conventional benchmark of $930\,$MeV}.
Parametrizations inside the window
describe self-bound matter, and hence stars made entirely of quarks; those
outside it describe matter that is unbound at low pressure, so that quarks can
only occupy the core of a hybrid star with a hadronic envelope. Because the
zero-pressure ground state may be two- or three-flavor, and because the
$p\!\to\!0$ limit must in addition be causal, the window is delimited by
several curves, which we now examine as functions of $\alpha$.

Figure~\ref{fig:1} shows the resulting windows for
$\alpha=0.67,\,1.00,\,1.14,\,1.50$. Four boundaries organize the plane:

\begin{itemize}
\item[(i)] $e^0_{ud}=930\,$MeV (black). Two-flavor matter contains no strange
quarks, so this curve is \emph{exactly} independent of $\alpha$ and provides a
fixed reference against which all the other displacements can be read.
\item[(ii)] $e^0_{uds}=930\,$MeV (red). Increasing $\alpha$ raises $M_s$ at a
given density and therefore raises $e^0_{uds}$, pushing the curve towards
smaller $C_{u,d}$ at fixed $a$.
\item[(iii)] $e^0_{ud}=e^0_{uds}$ (purple), which separates the parametrizations
whose $p=0$ ground state is two-flavor (to its left) from those whose ground
state is three-flavor (to its right). This is the boundary that moves the most:
its intersection with the $930\,$MeV curves shifts from $a\simeq1.1$ at
$\alpha=0.67$ to $a\simeq7.2$ at $\alpha=1.50$.
\item[(iv)] $c^0_s=c$ (gold), to the right of which the $p\to0$ limit is
acausal. This boundary is nearly vertical and, remarkably, nearly \emph{frozen}:
it moves only from $a\simeq5.29$ to $a\simeq5.69$ while $\alpha$ varies by more
than a factor of two.
\end{itemize}

The three energy boundaries meet at a single point, which we call the triple
point $(a_{\rm tp},C_{\rm tp})$: there $e^0_{ud}=e^0_{uds}=930\,$MeV. It is the
corner of the SQM domain, which is bounded on the left by the purple curve, from
above by the red curve, and on the right by the causality line. To compare the
position of that corner with the causality constraint we introduce
\begin{equation}
a_c(\alpha) \;\equiv\; a\big|_{c^0_s=c}\ \ \text{evaluated at}\ \
C_{u,d}=C_{\rm tp}(\alpha) ,
\label{eq:ac}
\end{equation}
that is, the abscissa of the causality boundary at the height of the triple
point. \RA{The SQM domain lies below the triple point, between the
$e^0_{ud}=e^0_{uds}$ curve and the causality line, and the horizontal gap
between these two curves shrinks monotonically as $C_{u,d}$ decreases (we
checked this numerically down to $C_{u,d}=2$). The whole domain is therefore
empty as soon as $a_{\rm tp}>a_c$.}
Table~\ref{tab:1} lists both quantities.

\begin{table}[tb]
\centering
\caption{Position of the triple point $(a_{\rm tp},C_{\rm tp})$ and value of the
causality boundary $a_c$ defined in Eq.~\eqref{eq:ac}, as functions of $\alpha$
\RA{(at $\kappa=0$)}.
The SQM domain closes when $a_{\rm tp}$ overtakes $a_c$. For
$\alpha\lesssim0.7$ the two $930\,$MeV curves are almost tangent over a wide
range of $C_{u,d}$ and the location of the triple point is correspondingly
ill-determined.}
\label{tab:1}
\begin{tabular}{ccccc}
\toprule
$\alpha$ & $a_{\rm tp}$ & $C_{\rm tp}$ & $a_c$ & \RA{$a_c-a_{\rm tp}$} \\
\midrule
0.67  & \RA{1.09}  & 138        & 5.29       & \RA{$+4.20$}  \\
0.75  & \RA{2.733} & \RA{82.7}  & \RA{5.358} & \RA{$+2.625$} \\
0.90  & \RA{4.074} & \RA{59.5}  & \RA{5.431} & \RA{$+1.357$} \\
1.00  & \RA{4.751} & \RA{50.9}  & \RA{5.477} & \RA{$+0.726$} \\
\RA{1.130} & \RA{5.506} & \RA{43.0} & \RA{5.536} & \RA{$+0.029$} \\
\RA{1.135} & \RA{5.533} & \RA{42.8} & \RA{5.538} & \RA{$+0.005$} \\
1.140 & \RA{5.560} & \RA{42.5}  & \RA{5.540} & \RA{$-0.020$} \\
1.30  & \RA{6.352} & \RA{35.8}  & \RA{5.610} & \RA{$-0.742$} \\
1.50  & \RA{7.206} & \RA{29.9}  & \RA{5.695} & \RA{$-1.511$} \\
\bottomrule
\end{tabular}
\end{table}

Two competing trends are now apparent. Raising $\alpha$ enlarges the region in
which the zero-pressure ground state has only two flavors, and correspondingly
displaces the triple point towards larger $a$ and smaller $C_{u,d}$. One might
expect that this simply trades SQM parametrizations for self-bound $ud$ ones.
The causality boundary, however, does not follow: it stays put. The SQM domain
is thus squeezed between a boundary that moves and one that does not, and it is
progressively extinguished.

The displacement originates in the balance that decides whether opening a new
Fermi sea pays off. Spreading a fixed baryon number over three flavors instead
of two lowers the Fermi momenta, and hence the kinetic energy of the light
sector; this is what makes strangeness attractive. The gain must be paid for
with the rest energy of the strange quasiparticles, which here is not the
current mass but the full effective mass
$M_s=m_s+\alpha C_{u,d}\,n_s^{-a/3}$. Raising $\alpha$ increases the medium
contribution to $M_s$, so the cost grows while the benefit does not: the balance
tips against the three-flavor phase, and the $e^0_{ud}=e^0_{uds}$ boundary is
pushed towards larger $a$, where the medium term decays faster with density and
the penalty is milder. The $c^0_s=c$ line, by contrast, probes the stiffness of
the mass functional --- controlled by $a$ --- and is therefore only weakly
sensitive to the relative coupling strength of the strange sector.

\section{Closing of the strange-matter window}
\label{sec:critical}

\begin{figure}[tb]
\centering
\includegraphics[width=0.92\columnwidth]{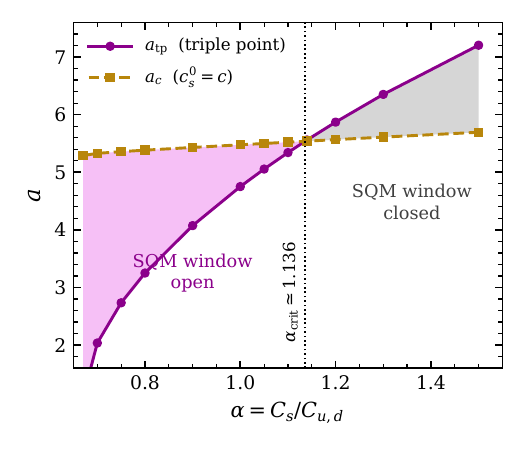}
\caption{Position of the triple point, $a_{\rm tp}$, and of the causality
boundary, $a_c$ of Eq.~\eqref{eq:ac}, as functions of $\alpha$. The SQM domain
exists only while $a_{\rm tp}<a_c$ (light magenta); the two curves cross at
\RA{$\alpha_{\rm crit}\simeq1.136$}, beyond which the window is closed (gray).}
\label{fig:2}
\end{figure}

The above competition is made quantitative in Fig.~\ref{fig:2}. The SQM domain
is non-empty if and only if the triple point lies on the causal side of the
$c^0_s=c$ line, i.e.\ $a_{\rm tp}(\alpha)<a_c(\alpha)$. Both quantities grow with
$\alpha$, but at very different rates: $a_{\rm tp}$ increases by a factor
$\sim7$ between $\alpha=0.67$ and $\alpha=1.5$, whereas $a_c$ increases by less
than $8\%$. The two curves therefore cross, and they do so at
\begin{equation}
\RA{1.135<\alpha_{\rm crit}<1.140 ,}
\label{eq:alpha_crit}
\end{equation}
\RA{with $a_{\rm tp}=a_c$ at $\alpha_{\rm crit}=1.136$. A direct scan of the
$(a,C_{u,d})$ plane for parametrizations with $e^0_{uds}<930\,$MeV,
$e^0_{uds}<e^0_{ud}$ and $c^0_s\le c$ confirms this bracket: a thin sliver
around $(a,C_{u,d})\simeq(5.54,42.6)$ survives at $\alpha=1.135$, and nothing at
$\alpha=1.140$.} Already at $\alpha=1$ the SQM domain is a narrow wedge
between \RA{$a\simeq4.75$} and the causality line (Fig.~\ref{fig:1}b), and by
$\alpha=1.14$ it has been squeezed shut. \RB{For $\kappa>0$ the electron
correction mentioned in Sec.~\ref{sec:model} lowers $e^0_{ud}$ and shifts the
closure to slightly smaller values, $\alpha_{\rm crit}\simeq1.11$ ($1.06$) for
$\kappa=0.3$ ($0.6$).}

The physical statement is therefore sharp. \emph{For
\RA{$\alpha>\alpha_{\rm crit}$} the
flavor-dependent QMDD model admits no causal, absolutely stable three-flavor
phase at zero pressure.} Every acceptable self-bound parametrization then has a
two-flavor ground state at $p=0$ and, since a new Fermi sea must eventually open
as the pressure increases, necessarily develops a first-order $ud\!\to\!uds$
transition at some finite pressure. Whether the conversion is realized inside a
compact star is a further, and logically independent, question: it requires the
transition density to be reached before the maximum-mass configuration, i.e.\
$p_c^{\max}\ge p_{\rm tr}$. For $\alpha=1$ this was found to hold over the whole
self-bound window~\citep{Teruya:2025tpx}, so that self-bound hybrid stars were
the generic outcome; in general, however, the transition may well be a feature
of the phase diagram of strongly interacting matter without ever appearing in
the stellar interior. Conversely, for $\alpha\lesssim0.7$ the two
$930\,$MeV curves become nearly tangent over an extended range of $C_{u,d}$ and
the SQM domain attains its maximal extension, recovering the traditional
strange-star picture.

Being nearly independent of $\alpha$, the $c^0_s=c$ boundary acts as a hard wall
that the displaced energy boundaries eventually hit. This answers, in the
negative, the question of whether arbitrarily large $\alpha$ can be used to
enlarge the useful parameter space: it cannot, because the space it opens lies
beyond the causal limit.

\section{Consequences for the $ud\!\to\!uds$ transition}
\label{sec:transition}

\begin{figure*}[tb]
\centering
\includegraphics[width=0.98\textwidth]{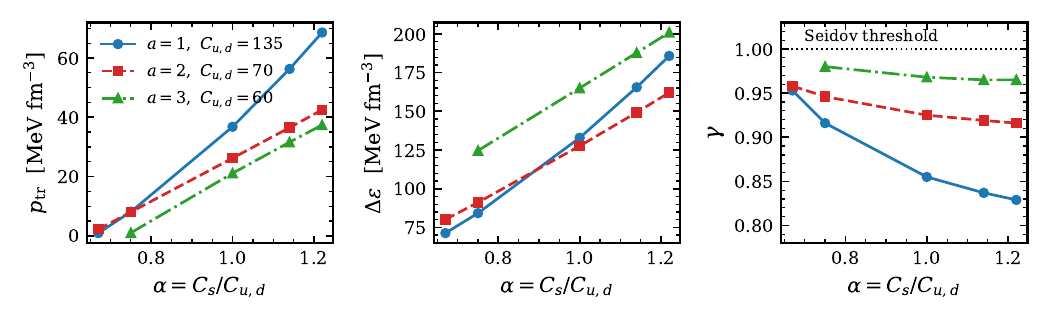}
\caption{Transition pressure $p_{\rm tr}$, energy-density jump
$\Delta\epsilon$, and Seidov diagnostic $\gamma$ of the first-order
$ud\!\to\!uds$ transition as functions of $\alpha$, for the three benchmark
parameter sets of Table~\ref{tab:2} at $\kappa=0$. Values for $\kappa>0$ follow
\RB{approximately} from the scaling law of Eq.~\eqref{eq:qscaling}; $\gamma$ is
\RB{nearly} invariant.}
\label{fig:3}
\end{figure*}

Whenever the zero-pressure ground state is two-flavor, a first-order
$ud\!\to\!uds$ conversion must take place at some finite pressure --- and above
$\alpha_{\rm crit}$ that is the situation of the entire causal self-bound
window. It is therefore natural to ask how the properties of that conversion
respond to $\alpha$. We locate it by equating the Gibbs free energies per baryon
of the two phases, and list the results in Table~\ref{tab:2} for the three
benchmark sets of~\cite{Teruya:2025tpx}, shown as diamonds in Fig.~\ref{fig:1};
Fig.~\ref{fig:3} displays them. The only entry missing from the table is
$(a,C_{u,d})=(3,60)$ at $\alpha=0.67$: as Fig.~\ref{fig:1}a shows, that point
then falls inside the (wide) SQM domain, its zero-pressure ground state is
already three-flavor, and consequently no $ud\!\to\!uds$ transition exists.

\begin{table}[tb]
\centering
\caption{First-order $ud\!\to\!uds$ transition for the benchmark parameter sets
at $\kappa=0$: transition pressure, energy densities on both sides, jump, and
Seidov diagnostic $\gamma$ of Eq.~\eqref{eq:seidov}. Pressures and energy
densities are in MeV\,fm$^{-3}$. Results for $\kappa>0$ follow
\RB{approximately} from Eq.~\eqref{eq:qscaling}.}
\label{tab:2}
\begin{tabular}{cccccc}
\toprule
$\alpha$ & $p_{\rm tr}$ & $\epsilon_{+}$ & $\epsilon_{-}$ & $\Delta\epsilon$ & $\gamma$ \\
\midrule
\multicolumn{6}{c}{$a=1$, $C_{u,d}=135$} \\
0.67 &  0.9 & 234.5 & 163.2 &  71.3 & 0.953 \\
0.75 &  8.1 & 279.4 & 195.2 &  84.2 & 0.916 \\
1.00 & 36.8 & 436.7 & 303.8 & 132.9 & 0.855 \\
1.14 & 56.3 & 537.0 & 371.5 & 165.5 & 0.837 \\
1.22 & 68.6 & 598.3 & 412.5 & 185.8 & 0.829 \\
\midrule
\multicolumn{6}{c}{$a=2$, $C_{u,d}=70$} \\
0.67 &  2.2 & 256.2 & 176.1 &  80.1 & 0.958 \\
0.75 &  8.0 & 281.4 & 190.3 &  91.1 & 0.946 \\
1.00 & 26.2 & 363.4 & 235.8 & 127.6 & 0.925 \\
1.14 & 36.5 & 411.3 & 262.0 & 149.3 & 0.919 \\
1.22 & 42.5 & 439.3 & 277.2 & 162.1 & 0.916 \\
\midrule
\multicolumn{6}{c}{$a=3$, $C_{u,d}=60$} \\
0.75 &  1.0 & 386.5 & 262.0 & 124.6 & 0.980 \\
1.00 & 21.1 & 462.5 & 297.5 & 165.0 & 0.968 \\
1.14 & 31.6 & 505.0 & 317.2 & 187.8 & 0.965 \\
1.22 & 37.4 & 529.3 & 328.4 & 200.9 & 0.965 \\
\bottomrule
\end{tabular}
\end{table}

Three features stand out.

\emph{(a) Monotonic growth.} Both $p_{\rm tr}$ and $\Delta\epsilon$ increase
monotonically with $\alpha$, and steeply so. For $a=1$, $C_{u,d}=135$ the
transition pressure grows from $0.9$ to $68.6\,$MeV\,fm$^{-3}$ when $\alpha$
goes from $0.67$ to $1.22$. This is the expected counterpart of the displacement
of the $e^0_{ud}=e^0_{uds}$ boundary: the more the strange sector is suppressed,
the higher the pressure required for the conversion and the larger its latent
heat. Should such a configuration be realized in a star, larger $\alpha$ would
therefore mean a smaller three-flavor core but a sharper density discontinuity
at the interface.

\emph{(b) \RB{Scaling} with the excluded volume.} The transition point is
determined by the crossing of the two Gibbs energies per baryon, and in the
excluded-volume construction $\mu_B$ is unaffected by $\kappa$ once the
thermodynamics is expressed in terms of the rescaled densities. Consequently the
transition occurs at the same $\mu_B$ for all $\kappa$. The pressure, the energy
density and the baryon number density, on the other hand, are volume densities
of quantities carried by a quark sector that occupies only the fraction $q$ of
the total volume, and are therefore all reduced by that same factor,
\begin{equation}
p_{\rm tr}(\kappa) = q\,p_{\rm tr}(0), \qquad
\epsilon_{\pm}(\kappa) = q\,\epsilon_{\pm}(0), \qquad q=1-\kappa ,
\label{eq:qscaling}
\end{equation}
whereas quantities defined per baryon, such as $\epsilon/n_B$, are left
unchanged. \RB{This is exact for the quark sector only: the electrons are not
rescaled and, through charge neutrality, make the equilibrium composition
weakly dependent on $\kappa$. For $(a,C_{u,d})=(1,135)$ at $\alpha=1$,
$p_{\rm tr}$ exceeds $q\,p_{\rm tr}(0)$ by $0.07\%$ ($0.26\%$) for $\kappa=0.3$
($0.6$), $\epsilon_\pm$ deviate by at most $0.15\%$, and $\gamma$ changes from
$0.8549$ to $0.8547$ ($0.8543$). The deviation grows with the electron fraction
of the $ud$ phase and reaches $3.7\%$ in $p_{\rm tr}$ for $(3,60)$ at
$\alpha=1$ and $\kappa=0.6$. In absolute terms it stays below
$0.35\,$MeV\,fm$^{-3}$ for all the entries of Table~\ref{tab:2}, so
Eq.~\eqref{eq:qscaling} is unreliable only where $p_{\rm tr}$ itself is of
order $1\,$MeV\,fm$^{-3}$.} It explains
the \RB{near} $\kappa$-independence of the Seidov diagnostic noted
\RE{by \citet{Teruya:2025tpx}} and it means that a single row per $(a,C_{u,d},\alpha)$
suffices to describe the whole $\kappa$ family \RB{to this accuracy}.

\emph{(c) Stability of the interface.} Seidov's criterion~\citep{Seidov1971}
\RC{concerns the star in which an infinitesimal core of the dense phase first
appears, assuming that a stable $ud$ sequence reaches $p_{\rm tr}$: that
configuration is unstable} if the energy-density discontinuity exceeds
$\epsilon_+/\epsilon_-=\tfrac32(1+p_{\rm tr}/\epsilon_-)$. Defining
\begin{equation}
\gamma \equiv \frac{\epsilon_{+}/\epsilon_{-}}
                   {\tfrac{3}{2}\left(1+p_{\rm tr}/\epsilon_{-}\right)} ,
\label{eq:seidov}
\end{equation}
we find $\gamma<1$ for every case examined. By Eq.~\eqref{eq:qscaling}, $\gamma$
is \RB{nearly} invariant under changes of $\kappa$. It does decrease with $\alpha$ --- from
$0.953$ to $0.829$ for the $a=1$ family --- so larger $\alpha$ moves the
configurations \emph{away} from the Seidov threshold rather than towards it,
even though $\Delta\epsilon$ itself grows. \RC{The onset of the $uds$ core is
therefore not destabilizing. How far the hybrid branch extends, and whether
$p_{\rm tr}$ is reached at all before the maximum mass, can only be decided
with full general-relativistic sequences such as those computed}
\RE{by \citet{Teruya:2025tpx}} for $\alpha=1$.

\section{Conclusions}
\label{sec:conclusions}

We have extended the flavor-dependent QMDD model by allowing the strange quark
to carry its own coupling in the mass functional, $C_s=\alpha\,C_{u,d}$, and we
have mapped the zero-pressure parameter space as a function of $\alpha$. Our
main conclusions are:

\begin{itemize}
\item[(i)] The $e^0_{ud}=930\,$MeV boundary is exactly $\alpha$-independent,
while the three-flavor boundaries move systematically with $\alpha$. Increasing
$\alpha$ enlarges the self-bound two-flavor domain at the expense of strange
quark matter and displaces the triple point from $(a,C_{u,d})\simeq(1.1,138)$ at
$\alpha=0.67$ to $\simeq(7.2,30)$ at $\alpha=1.5$.

\item[(ii)] The causality boundary $c^0_s=c$ is nearly frozen at
$a\simeq5.3$--$5.7$. As a result the strange-matter window closes at
\RA{$\alpha_{\rm crit}\simeq1.136$} \RB{(at $\kappa=0$; slightly earlier for
$\kappa>0$)}: above this value no causal, absolutely stable
three-flavor phase exists at $p=0$, so that self-bound matter in this model is
necessarily two-flavor at zero pressure and must undergo a first-order
$ud\!\to\!uds$ transition at some higher pressure. Whether that transition is
reached inside a star depends on whether the transition density lies below the
central density of the maximum-mass configuration. Enlarging $\alpha$ beyond
$\alpha_{\rm crit}$ opens no new usable parameter space.

\item[(iii)] The transition pressure and the energy-density jump grow
monotonically with $\alpha$, whereas the Seidov diagnostic decreases and remains
below unity\RC{; provided a stable $ud$ sequence reaches $p_{\rm tr}$, the
onset of the $uds$ core is therefore not destabilizing}.

\item[(iv)] The excluded-volume parameter acts as an overall rescaling of
the transition quantities, $p_{\rm tr},\epsilon_\pm \propto (1-\kappa)$, leaving
the Seidov diagnostic \RB{nearly} invariant. \RB{The rescaling is exact for the
quark sector; the unscaled electrons add corrections of at most a few percent,
except where $p_{\rm tr}$ is very small.}
\end{itemize}

A natural continuation is to propagate the present parameter-space analysis into
stellar sequences, in order to quantify how $\alpha$ redistributes mass, radius
and tidal deformability along the self-bound hybrid branch.

\section*{Acknowledgements}

A.G.G. acknowledges the financial support from CONICET (Argentina) under Grant
No. PIP 2022-2024 GI - 11220210100150CO, and from the National University of La
Plata (Argentina), Project No. X960. G.L. acknowledges the financial support
from the Brazilian agency CNPq (grant 316844/2021-7).

\makeatletter
\renewcommand{\doi}[1]{\RG{\urlstyle{same}\@ifundefined{href}{\url{#1}}{\href{https://doi.org/#1}{\nolinkurl{#1}}}}}
\makeatother
\bibliography{references}

\end{document}